\documentclass{article}
\usepackage[margin=2.5cm]{geometry}
\usepackage[T1]{fontenc}
\usepackage[obeyspaces]{url}
\usepackage{harvard}
\usepackage{xcolor}
\usepackage{comment}
\usepackage{fancyhdr}
\begin{document}

\title{Database Reconstruction
Is Not So Easy and Is Different from Reidentification\thanks{
Support from the European Commission (projects H2020-871042
``SoBigData++'' and H2020-101006879 ``MobiDataLab'') and 
the Government of Catalonia
(ICREA Acad\`emia Prize to J. Domingo-Ferrer)
is gratefully acknowledged. 
}}

\author{
Krishnamurty Muralidhar$^1$ and Josep Domingo-Ferrer$^2$\\
{\small $^1$ University of Oklahoma}\\ 
{\small Price College of Business}\\
{\small Dept. of Marketing and Supply Chain Management}\\
{\small 307 West Brooks, Adams Hall Room 10}\\
{\small Norman, OK 73019, U.S.A.}\\
{\small {\tt krishm@ou.edu}}\\
{\small $^2$ Universitat Rovira i Virgili}\\
{\small Dept. of Computer Engineering and Mathematics}\\
{\small CYBERCAT-Center for Cybersecurity Research of Catalonia}\\
{\small Av. Pa\"{\i}sos Catalans 26}\\
{\small 43007 Tarragona, Catalonia}\\
{\small {\tt josep.domingo@urv.cat}}}

\maketitle

\begin{abstract}
In recent years, it has been claimed \textcolor{black}{that releasing} 
 accurate statistical information on a database is likely to allow its complete reconstruction. Differential privacy has been suggested as 
%JOSEP2208. Rev. 2. Changed "the only way" to "the gold standard".
the \textcolor{black}{appropriate methodology} 
to prevent these attacks. These claims have recently been taken very seriously by the U.S. Census Bureau and led them to adopt differential privacy for releasing U.S. Census data. This in turn has caused consternation among users of the Census data due to the lack of accuracy of the protected outputs. It has also brought legal action against the U.S. Department of Commerce. In this paper, we trace the origins of the claim that releasing information on a database automatically makes it \textcolor{black}{vulnerable to being
exposed by} reconstruction attacks and we show that this claim is, in fact, incorrect. We also show that reconstruction can be averted by properly using traditional 
statistical disclosure control (SDC) techniques. We further show that the geographic level at which exact counts are released is even more relevant to protection than the actual SDC method employed. Finally, we caution against confusing reconstruction and reidentification: using the quality of reconstruction as a metric of reidentification results in exaggerated reidentification risk figures.\\

{\bf Keywords:} Database privacy, Database reconstruction, Statistical
disclosure control, Differential privacy.
\end{abstract}

\section{Introduction}
\label{introduction}

Database reconstruction seems to be the nemesis of
official statistics and statistical data release as they
have been known so far. According to the U.S. Census Bureau's
Chief Scientist:
\begin{quotation}
This [Dinur and Nissim's database reconstruction] theorem
is the death knell for public-use detailed 
tabulations and microdata sets as they have been 
traditionally prepared. \cite{abowd17,abowd19}
%The database reconstruction theorem is the
%death knell for traditional data publication
%systems from confidential sources.\citeasnoun{abowd19}
\end{quotation}

Whenever a database contains personal information on a set of respondents,
%JOSEP2208. Rev. 2. Changed. 
data protection legislation {may require} 
the organization in charge of a database, called {\em controller} 
in the European legal parlance~\cite{GDPR}, to take steps to protect
respondent privacy.
%JOSEP2208. Rev. 4. Added.  
\textcolor{black}{S}tatistical disclosure control 
(SDC, \cite{dalenius,hundepool}) is a discipline that provides 
methods to this end. SDC methods operate by 
masking, {\em i.e.} altering, the data to be protected; masking can be
based on data perturbation, on reduction of detail or even 
on generating synthetic data that preserve some of the statistics
of the original data. 
 Depending on when masking is applied, the 
SDC literature distinguishes among \textcolor{black}{{\em local}} protection
(where 
data are masked by respondents themselves before being collected),
{\em input} protection (data are masked by the controller after
collecting them and all subsequent queries are answered based
on the masked data) and {\em output} protection (queries
are computed on the true original respondent data and the 
query outputs are masked before being released).

%JOSEP. Citro comment
\textcolor{black}{
The process of forming a database can take place at a certain point
in time or be continuous during all the lifetime of the database.
In the former case, the database is said to be static, whereas in the latter
case it is said to be dynamic. In static
databases, the data are first collected and
then they are structured to form the database. This yields
a ``frozen'' database which is subsequently used
to answer any queries.
%JOSEP2208. Rev. 2. Changed "constantly" to "periodically"
In contrast, dynamic databases contain data that are
{periodically}
changing, with new records or even attributes being added and removed
over time. Organizational and corporate databases
({\em e.g.}, containing data on customers, orders, etc.) are usually
dynamic. Static databases are typical in data collected for research
and certain data gathered by government agencies. Obviously, not
all government data are static, but in many situations they are.}
%Our focus in what follows will be on static databases.

Output protection is the most convenient option for dynamic databases, 
as it avoids having to create masked versions of the underlying 
changing \textcolor{black}{data.} 
%However, output protection is inherently more disclosive 
%than input or \textcolor{black}{local} protection, because the former
%approach attempts to protect answers that were
%computed on the original respondent data, whereas in the 
%latter approaches the answers are computed on modified data. 
\textcolor{black}{Whatever} 
the type of protection, the level of protection achieved
depends on the extent to which the data have been 
modified \textcolor{black}{\cite{dwork06}}. In general,
the greater the modification, the greater the protection but the lesser
the accuracy and hence the utility of the protected data. 
%JOSEP2208. Rev. 2. Suppressed sentence.
%The amount of modification required to attain a certain level of protection
%is also dependent on the characteristics of the data: the less diverse
%the data ({\em i.e.} the more similar
%the records of respondents), 
%the less modification is needed. 
In particular, if the 
original data contain outliers or unique records, greater modification
may be necessary.
For further discussions on SDC and the privacy-utility trade-off,
see \textcolor{black}{\citeasnoun{traub84}}, 
\citeasnoun{Adam89}, \textcolor{black}{\citeasnoun{duncan01}}, 
and \citeasnoun{hundepool}.

The possibility of mounting reconstruction attacks
has been known for decades, and a formal theory of reconstruction
attacks was developed by Dinur and Nissim 
twenty years ago~\cite{dinur03}.
%JOSEP2208. Rev 2. "claimed" to "stated".
U.S. Census Bureau methodologists recently {stated}
that such  attacks are no longer just a theoretical possibility,
but a practical danger. Hence, they advocate using
differential privacy~\cite{dwork06,dworkroth} ---DP in what follows--- 
to protect the statistical
outputs of the U.S. 2020 Decennial Census as a way
to thwart reconstruction of the underlying microdata.
The decision to use DP motivated a lawsuit from the 
State of Alabama against the U.S. Department of Commerce, basically
arguing utility loss \textcolor{black}{(and a delay in the data release)}~\cite{Alabama}. This lawsuit was backed
by 16 other states~\cite{Associated}, but it was recently
rejected by the judges~\cite{percival}, on the grounds that no damage
to Alabama will be provable until the DP Census 
data are made available.

In April 2021, the Census Bureau  published
a version of the 2010 Decennial Census using their new DP-based
methodology, called DAS. After studying that version, several users 
have expressed their concerns about the utility loss caused by 
%JOSEP2208. Rev. 2. Added reference to Hotz and ons
DAS~\cite{kenny21,ruggles21,hotz,ons,apnews}.

Since using \textcolor{black}{a DP-based methodology 
to prevent reconstruction} 
%JOSEP2208. Rev. 2. Removed "at best"
is controversial in terms of utility, it remains
extremely relevant to examine the real danger of reconstruction
attacks and the extent to which such attacks can be warded off by DP 
or other methods at a reasonable utility cost.

\subsection*{Contribution and plan of this paper}

%JOSEP. Citro comment.
In this paper, we reassess the risk of the original data
being reconstructed by an attacker based on the protected
query outputs. We first give background on reconstruction 
attacks \textcolor{black}{(Section~\ref{reconstruct})}. We then examine 
the protection that DP can offer against
reconstruction \textcolor{black}{(Section~\ref{dpreconstruct})}.
%JOSEP2208. Rev. 2. Suppressed.
% and we compare it against the protection
%that can be obtained using a much older SDC method such
%as randomized response (RR). 
%JOSEP2208. Rev. 1. Suppressed.
%We then move to a case study drawn
%from an article   
%by U.S. Census Bureau methodologists where they 
%claim that traditional SDC methods (specifically cell suppression) 
%do not protect against reconstruction attasks and they propose 
%to use differential privacy instead; we show that if cell suppression
%had been correctly applied in their case study, 
%it would protect against reconstruction 
%with much less utility damage than DP. 
After that \textcolor{black}{(Section~\ref{geography})}, 
we discuss the critical relevance 
for reconstruction of the geographic level
at which exact counts are reported\footnote{{In fact, the Census Bureau 
has released the production run of the counts needed for redistricting
(\url{https://www.census.gov/newsroom/press-kits/2021/2020-census-redistricting.html}).}}.
In Section~\ref{linkage} we highlight the 
differences
between reconstruction and reidentification: using
the quality of reconstruction as a measure of reidentification
risk results in exaggerated reidentification risk.
%Section~\ref{reconstruct} recalls reconstruction attacks.
%Section~\ref{dpreconstruct} evaluates the protection offered by 
%DP against reconstruction attacks and compares it with what
%can be done using randomized response (RR). 
%%JOSEP2208. Rev. 1. Deleted.
%Section~\ref{dpcensus} analyzes  
%the aforementioned case study proposed by U.S. Census Bureau methodologists. 
%Section~\ref{geography} highlights that the geography 
%size for which counts are preserved can be more relevant for reconstruction
%than the SDC method used to protect data.
%In Section~\ref{linkage} we caution against using
%the quality of reconstruction as a metric for reidentification risk.
Conclusions and future research lines are summarized 
in Section~\ref{conclusion}.

\section{Reconstruction attacks}
\label{reconstruct}

Reconstruction attacks have been known for a long time 
in the literature. In their pioneering work,
%Denning and Schl\"{o}rer~
\citeasnoun{denn80} showed that a poorly 
designed database query answering system based on {\em output
perturbation} can easily lead to disclosure of some or even all
of the database records. The tool they used was the tracker attack,
a carefully crafted sequence of queries aimed at isolating 
and disclosing specific records. 

A more formal analysis of the ability to reconstruct the contents
of a database using only the outputs of queries was formulated
by 
%Dinur and Nissim
\citeasnoun{dinur03}, hereafter DN. 
The main step forward is their discovery that the attacker
does not even need to be careful when constructing her sequence
of queries. 
The authors assume
the database is an $n$-bit string, that is, it contains records
each of which takes value 0 or 1. They further assume all queries
to be of the form ``How many records in this subset are 0's?'' or
``How many records in this subset are 1's?''. In their setting,
the response to every query is computed as the true query 
answer plus an error $E$ bounded in an interval $[-B,B]$ for some
$B>0$. Thus, it is clear that DN assume that 
protection of query outputs
is performed via {\em output perturbation} {and that the error 
is strictly bounded}. 

A database reconstruction, according to DN, is a record-by-record
%JOSEP2208. Rev. 2. Added.
reconstruction of the original values {such that
the {\em distance between the reconstructed values and the original
values is within specific accuracy bounds}.}
Thus, DN's attacks are aimed
at inferring the value of {\em each record in the original 
database} with 
a high level of accuracy. They consider two different attackers depending 
on their computational power:
\begin{itemize}
\item {\em Exponential attacker.} This attacker is able to issue all
possible queries. In practice, such an adversary is only realistic
for {\em small databases}, because, say for $n\geq 100$, it would take
years or decades to issue all possible queries, even with the fastest
computers available. To protect against such an attacker, the output
of any query is modified by adding random noise in $[-B,B]$. If
the differences between the query responses obtained on the target
original database and the corresponding query responses obtained on 
a specific candidate database are within $B$, then the candidate
database represents a reconstruction of the original database. DN
show that, in this case, the candidate database is within distance
$4B$ of the target original database, where both databases
are taken as binary $n$-vectors. Thus, unless the value $B$ is relatively
large, the candidate database is a {\em good} reconstruction of the original
database. DN proved that, in order to prevent such a good reconstruction, 
$B$ must be non-negligible compared to $n$, that is, $B$ must be $O(n)$. 
\item {\em Polynomial attacker.} This attacker issues a number of 
queries that is polynomial in $n$, which is feasible 
for large databases. The database protects the query
outputs by adding random noise in $[-B,B]$. Using these protected 
query outputs, the attacker solves a linear programming problem
to reconstruct the database. DN show that, with high probability, 
the reconstructed database is close to the original database as long
as $B$ is within $o(\sqrt{n})$ where the ``little o'' notation means
much smaller than $\sqrt{n}$ as $n$ grows. Hence, to achieve protection,
$B$ must be $O(\sqrt{n})$ (with ``big O'') for a non-negligible proportion
of queries.
\end{itemize}

Thus, DN conclude that, unless the noise added to query outputs
is commensurate to the size of the database
($O(n)$ for an exponential adversary and $O(\sqrt{n})$ for a polynomial
adversary), the attacker is able to recreate the database. Is a noise level
at least $O(\sqrt{n})$ realistic? Consider a database of size
$n=$1,000,000. What is being required is that the answers to
a {\em non-negligible proportion} of queries differ from the 
corresponding true answers by about 1,000. 
Note that this noise does not have to be applied to all queries.
%Note that 1,000 is the maximum
%noise and that the noise bounded in $[-1,000, 1,000]$ 
%does not have to be applied to all queries. 
Furthermore, a perturbation
of about 1,000 is relatively small compared to the size of the database
and to queries that may involve several hundred thousand records. 
Thus, the noise level required to protect against a polynomial
adversary seems affordable in many situations.

Without question, 
%the DN paper~
%\citeasnoun{dinur03} 
DN give very relevant
insights into database reconstruction using only responses 
to queries. The authors give a theoretical framework that explains
the reconstruction risk as a function of the adversary's computational
power and the noise applied to query outputs. 
Yet, {\em providing a theoretical framework for database reconstruction
does not mean that every database can be reconstructed.}

For one thing, {\em the results by DN apply only to output perturbation,
but not to \textcolor{black}{local} or input protection.}
{This is explicitly acknowledged by DN when
they mention the ``CD Model'':
\begin{quotation}
{\bf The CD Model.} The database algorithm above essentially
creates a `private' version of the database $d'$, and then
answers queries using $d'$. Note that a user may retrieve the entire
content of $d'$ by querying $q_i =\{i\}$ for $1\leq i \leq n$, after
which she may answer all her other queries by herself. This result
indicates that it is in some cases possible to achieve privacy
in a {\em CD model}, where users get a `private' version of the
database (written on a CD), which they may manipulate
(say, without being restricted to statistical queries).
\end{quotation}}
%JOSEP2208. Rev. 1 and Rev. 5.
Specifically, if \textcolor{black}{local} or input masking are implemented, 
the responses to all queries 
are based on the masked database. Hence, {\em for \textcolor{black}{local} 
or input perturbation,
the DN framework can only reconstruct the masked database.}  
Now, if the \textcolor{black}{local} or input masking are configured to adequately
protect the original database ({\em e.g.}, using RR at the 
\textcolor{black}{respondent's} or 
  microdata SDC methods described in \citeasnoun{hundepool}),
reconstructing the masked database should not entail
disclosure of sensitive information.
%As mentioned 
%In practice, input-masked data, a.k.a. SDC-protected microdata, 
%are considered safe enough to be released in their entirety or partially
%({\em e.g.} a random sample) to users for them to perform their
%own analyses. 
%Hence, it is not a major concern if an attacker reconstructs
%a source or input-protected database. 

%In fact, DN acknowledge that their 
%reconstruction does not apply to input-protected data, 
%when they mention the ``CD Model'':
%\begin{quotation}
%{\bf The CD Model.} The database algorithm above essentially
%creates a `private' version of the database $d'$, and then 
%answers queries using $d'$. Note that a user may retrieve the entire
%content of $d'$ by querying $q_i =\{i\}$ for $1\leq i \leq n$, after
%which she may answer all her other queries by herself. This result
%indicates that it is in some cases possible to achieve privacy
%in a {\em CD model}, where users get a `private' version of the 
%database (written on a CD), which they may manipulate
%(say, without being restricted to statistical queries).
%\end{quotation}
%DN's ``CD Model'' exactly describes SDC-protected microdata released
%to users for them to perform their own analyses. 
%Given that query outputs computed on source or input-protected
%microdata allow at best reconstructing the protected microdata
%(but not the original microdata), maybe the sky is not falling
%after all. 

\section{The performance of differential privacy against reconstruction}
\label{dpreconstruct}

In~\citeasnoun{dwork11} and \citeasnoun{garfinkel19}, the purported solution to the 
reconstruction vulnerability of output-protected data is 
differential privacy (DP). 
%We next briefly recall DP.
%\subsection{Background on differential privacy}
DP was introduced by \citeasnoun{dwork06} as a
framework for quantifying the disclosure risk associated with answering
queries based on a confidential database. Assume an adversary
 submits a query to the database and obtains a query response $R$. 
$\epsilon$-DP requires that, given
two databases $D$ and $D'$ that differ in one record, and for all
subsets $S$ of the space of query responses
\begin{equation}
\label{epsilon}
 \Pr(R \in S| D \mbox{ is used}) \leq e^{\epsilon} \times 
\Pr(R \in S| D'\mbox{ is used}).
\end{equation}
Essentially, DP requires that, by observing $R$, 
it must be indistinguishable within a factor
$e^{\epsilon}$ whether the database $D$ or the database $D'$
are being used.
When $\epsilon=0$, this requirement implies that the database in use
must be completely indistinguishable when observing $R$. 
 In this case, the value of the record differing between $D$ and $D'$
stays completely confidential in spite of $R$ 
being returned to the adversary. The value $\epsilon$ is usually
called ``privacy budget'' and it should be small for 
the privacy condition of Expression (\ref{epsilon}) to be meaningful: 
\citeasnoun{dwork11} recommended
$\epsilon$ to be ``say, 0.01, 0.1, or in some cases, $\ln 2$
or $\ln 3$.'' 

%The DP definition recalled above can be viewed as a variation of 
%the one used in 2003 by DN when they describe a two-phased adversary
%who attemps to predict the value of a target record in a database
%of $n$ records given 
%information on the remaining $n-1$ records. But there is 
%an important different between DN's definition and DP: whereas
%DN's adversary focuses on a specific target record,
%DP requires protecting {\em any} single record, even those that
%are not in the database.     

A well-known property of DP is sequential composition: if $k$ 
queries are individually answered with privacy levels $\epsilon_1$, 
$\epsilon_2$, $\ldots$, $\epsilon_k$, respectively, 
the extant privacy level after answering all $k$ queries 
is $\epsilon_1 + \epsilon_2 + \cdots + \epsilon_k$.

A relaxation of DP called $(\epsilon,\delta)$-DP has also been 
proposed and is defined as
\begin{equation}
\label{delta}
\Pr(R \in S| D\mbox{ is used}) \leq e^{\epsilon} \times 
\Pr(R \in S| D'\mbox{ is used}) + \delta,
\end{equation}
where $\delta$ is the relaxation parameter. The value of $\delta$
is often interpreted to imply that $\epsilon$-DP is satisfied
with probability $1 - \delta$. But a closer comparison between
Expressions (\ref{epsilon}) and (\ref{delta}) suffices to realize that 
$(\epsilon,\delta)$-DP can hold without $\epsilon$-DP being satisfied
for {\em any} query.

The usual procedure to achieve differential privacy 
is to return a query answer $R$ that 
consists in the query result computed on the original 
data plus Laplace-distributed noise. The smaller the value 
of $\epsilon$ and the more sensitive the query ({\em i.e.} the larger
the potential change of the query result due to the change of a single
record), the greater the amount of noise required.

%\subsection{Differential privacy and reconstruction}

\citeasnoun{dwork11} seems to suggest that 
the $O(\sqrt{n})$ 
accuracy provided by randomized response~\cite{warner65}\footnote{{Even though RR can satisfy the DP requirements~\cite{dwork11},
RR was proposed in 1965, more than a decade before the birth of the SDC
discipline and four decades before DP. In fact, RR  
has other properties beyond DP, such as allowing an estimation 
of the original distribution based on the randomized distribution.}}
can 
be outperformed by a differentially private procedure, when she writes:
\begin{quotation}
Suppose $n$ respondents each employ randomized response independently,
but using coins of known, fixed, bias. Then, given the randomized
data, by the properties of the binomial distribution the analyst
can approximate the true answer to the question ``How many respondents
have value $b$?'' to within an expected error on the order of $O(\sqrt{n})$.
As we will see, it is possible to do much better---obtaining constant expected
error, independent of $n$.
\end{quotation}
Yet, achieving constant error independent of $n$ clashes with 
the requirement of \citeasnoun{dinur03} according to which, to 
prevent a (polynomial) adversary from being able to reconstruct
a database based on query outputs, noise at least $O(\sqrt{n})$ is 
needed for a non-negligible proportion of queries. 
As Dwork acknowledges above, achieving $O(\sqrt{n})$ noise is precisely
what randomized response does. 

Hence, if a differentially private
procedure offers constant error independent of $n$,
it cannot protect against reconstruction according to DN. 
In fact, if a DN-adversary is allowed to submit a polynomial 
number of queries, say $m=O(n)$, 
%JOSEP2208. Rev. 1. Changed. 
{sequential composition applies,
because in general queries may be on overlapping sets of individuals.}
%the above-mentioned sequential composition
%property of $\epsilon$-DP requires 
%that 
{Thus,} the total privacy budget 
$\epsilon$ must be split into chunks of $\epsilon/m$ per query.
Hence, since the noise applied to each query answer is inversely
proportional to its privacy budget, 
{for the Laplace mechanism 
the standard deviation of the noise} is directly
proportional to $m/\epsilon$ and therefore $O(n)$.
To summarize, if $\epsilon$-DP is correctly applied, it protects
against reconstruction because it uses $O(n)$ noise, considerably
more noise than randomized response.
%JOSEP2208. Krish. Added.
{Therefore, $\epsilon$-DP mechanisms are likely
to over-protect outputs of increasing complexity, as noted
in~\citeasnoun{bach}.}

%JOSEP2208. Rev. 1. Removed section.

\section{The relevance of geography and policy decisions for reconstruction}
\label{geography}

%\textcolor{black}{\em Krish, can you please send me the text on 
%what you suggested to add to this section?\\
%"In this section, in addition to the current discussion, discuss the use of Swapping versus DP in the context of the Census:\\
%     *   Show how Swapping is implemented,\\
%     *   Show how DP is implemented,\\
%     *   Show that Swapping is "exchange" of records between cells whereas DP is "deletion/addition" (add records to some cells and delete records in some cells),\\
%     *   Discuss how in terms of reidentification what matters is the CHANGE IN THE FREQUENCY count. Discuss that swapping can control this (since it is a \% change) whereas DP is independent of the cell count and could potentially be more disclosive.}

\textcolor{black}{U.S. law\footnote{\textcolor{black}{\url{https://www.census.gov/about/adrm/linkage/about/authority.html}}}} requires that
the Census Bureau not 
``make any publication whereby the data furnished by any particular
establishment or individual [...] can be identified.''
Any individual with unique characteristics at the lowest geographic
level at which the tables are released 
%JOSEP. Citro comment.
\textcolor{black}{is at risk of reidentification}; that is,
any cell count of 1 is exposed to 
reidentification.
%\footnote{The U.S. Census Bureau's Chief Scientist
%also mentions this point in his court declaration
%(see \citeasnoun{abowd21}, p. 65, and \citeasnoun{abowdadd21}, p. 11).
%He proposes to aggregate at higher geographic level: whereas
%this is done when using DP for the Census 2010, this is not done
%when using swapping. A fair comparison of DP and swapping requires
%a comparable level of aggregation.}.

One of the key reasons for implementing DP in the Census context 
is the claim that the 
swapping approach used to protect previous decennial censuses
was ineffective against reconstruction and reidentification.
\textcolor{black}{
\citeasnoun{garfinkel19} provide a simple hypothetical example
using primary and secondary suppression to highlight the danger
of reconstruction.
It has been shown \cite{cacmletter,unece21}
 that even this reconstruction} 
 would have been
infeasible if primary and secondary suppression had been 
 applied in the correct way ({\em e.g.}, as described in Census methodology
documentation~\cite{census20} and related SDC 
literature~\cite{antal17,unece15}).
It remains however true that publishing statistics at a detailed
geographic level may facilitate reconstruction. We examine
this issue in what follows.

\subsection{The impact of geography}
\label{geo}

If statistics are released in small geographies, reconstruction
\textcolor{black}{can be performed using simple arithmetic.}
In fact, no matter whether swapping or DP is used as an SDC
approach to  protect tables, if total 
counts are exactly preserved at a small geographic
level, then reconstruction 
is {feasible (see \citeasnoun{abowdhawes22}, p. 8)}. 
The ease of reconstruction greatly depends
on how small are the geographic areas for which exact counts are 
reported. 
More precisely, in the comparison between swapping and DP
on the Census 2010 data conducted by the U.S. Census Bureau:
\begin{itemize}
\item When implementing swapping on the 2010 Census, 
total population and voting age 
counts were held invariant (exactly reported) at the block 
level (\citeasnoun{abowd21}, p. 12). 
%Releasing exact count data at the block level 
%can be viewed as releasing a frequency table 
%{\em Occupied\_Block\_ID} $\times$ {\em Gender} $\times$ {\em Age} $\times$ 
%${\em Race} $\times$ {\em Ethnicity}. This table has 
%6,207,027 $\times$ 2 $\times$ 103 $\times$ 63 $\times$ 2 $=$ 161,109,592,812 cells.
%Since the total number of people in the 2010 Census (308,745,538) 
%is smaller than the total number of cells by a factor of 500, 
%the aforementioned table 
%is a sparse matrix with a high proportion of cell counts 
%equal to 1. 
\item In contrast, when implementing differential privacy
on the 2010 Census, only the state-level population was held invariant. 
\textcolor{black}{Note that in 2017, block-level exact
counts had been promised: ``By agreement 
with the Department of Justice, the Census Bureau will provide
exact counts at the Census block level [...]''~\cite{census17}.}
\end{itemize}

{Now, there are over 6 million blocks versus only \textcolor{black}{50 states
plus the District of Columbia.} 
As a result, swapping was far more constrained than
DP and, as a result, more disclosive.}
{\em Eliminating block-level constraints \textcolor{black}{
(preserving total population and voting-age population counts)}
 for swapping might 
%JOSEP. Citro comment.
\textcolor{black}{  
put the privacy protection afforded by swapping
on the same footing as DP.}} A fair comparison 
between swapping and DP would require the 
Census Bureau to report the results of their reconstruction 
attacks applied to both swapped data and DP-protected data from
 the 2010 Decennial Census when exact counts are preserved
%JOSEP2208. Rev. 5. Added footnote. 
at the same geographic level. This would allow comparing the protection
and the utility provided by both approaches; in particular,
it would be interesting to see the extent to which reconstruction
on DP-protected data can be performed in the same way described for 
swapping (\citeasnoun{garfinkel19}, p. 34).

Although the
Census Bureau claims to have performed a comparative analysis of DP 
against swapping and suppression, no specific comparative results are
available. Only the following statement is provided: ``to achieve
the necessary level of privacy protection, both enhanced data
swapping and suppression had severely deleterious effects on data
quality and availability'' (\citeasnoun{abowd21}, p. 25).
%Those ``deleterious effects'' were never quantified nor were
%the parameters used in the alternative data swapping or suppression models.

Another concern is that even relaxing from exact count 
preservation to consistent count preservation at several
geographic levels is problematic under DP.
According to~\citeasnoun{garfinkelPETS} (p. 59),
noise can be added in all geographic levels 
of the Census 2020 as long as consistency is maintained.
To ensure this consistency, the DAS methodology developed by 
the Census based on DP 
involves several postprocessing steps~\cite{kenny21}\textcolor{black}{.}
% This postprocessing is not part of DP
%and in fact, it may break the DP privacy guarantees and facilitate
%reconstruction in the case of small geographies\footnote{{As argued in~\citeasnoun{gong},
%any invariant constraints must be considered in the design of the DP
%mechanism (in what is called congeniality), because waiting to
%a postprocessing phase to enforce them may undermine DP.}}.

\subsection{The impact of the privacy budget on DP}

If DP is advocated as a replacement of traditional SDC
methods, the privacy budget $\epsilon$ should be specified, as
enjoined in~\citeasnoun{dwork19}. Taking 
\textcolor{black}{a very small $\epsilon$} entails
unaffordable utility loss, but taking $\epsilon$ very large
entails very little noise addition and
offers little to no protection against reconstruction, let alone
reidentification~\cite{domingo-ferrer21}.

In fact, recent U.S. Census documents
mention $\epsilon$ values as high as 19.61 \textcolor{black}{in 2021~\cite{census21} and 39.907 
in 2022~\cite{census22}. Let us take the 2021 $\epsilon$ value to illustrate
how little privacy it achieves (the 2022 value still achieves less).}
We offer two different views that lead
to similar conclusions:
\begin{itemize}
\item First, we use the connection between DP and randomized
response~\cite{wang16}. Consider RR for a binary attribute,
so that the reported randomized answer is equivalent to the true
answer with probability $p\geq 0.5$ and different with probability
$1-p$. Then, for any $\epsilon$, the disclosure risk incurred
by $\epsilon$-DP is the same incurred by RR when
$p= \exp(\epsilon)/(1 + \exp(\epsilon))$.  Specifically,
\textcolor{black}{$\epsilon = 19.61$ translates to binary RR with
$p=0.99999999696$}, that is,
to RR reporting the original value with probability practically 1,
which basically amounts to no disclosure protection being offered.
\item Alternatively, we take the Dinur 
and Nissim perspective. For illustrative 
purposes, assume that the noise added is sampled from a 
Laplace distribution. For \textcolor{black}{$\epsilon=19.61$, the noise is bounded
in the range $[-1,1]$ with probability higher than $0.999999997$,
and it is bounded in the range $[-0.5,0.5]$ with probability higher
than $0.999944825$}. Adding this level of noise would violate the DN
requirement that the noise should be $O(\sqrt{n})$ and would allow
accurate reconstruction of the data.
\end{itemize}
Worse yet, even with high values of $\epsilon$, the utility of the
released DP data can be very low in some cases, as
noted in~\citeasnoun{vanriper20}, \citeasnoun{ruggles21} 
and \citeasnoun{kenny21}.
This is 
\textcolor{black}{due to: (a) sequential composition, which requires splitting the privacy budget
among all released outputs that are not independent of each other
(for instance, among
all the cells of a table, among different geographic levels,
among queries related to each other, etc.); and (b) post-processing with 
the Census's TopDown Algorithm (TDA), which is required in order to publish
data that are consistent, integral and non-negative.}

The real protection and the
utility loss of the high values of $\epsilon$ being proposed
should be compared
to those achievable using traditional SDC methods
({\em e.g.}, those employed in the 2010 Decennial Census) under
the same invariance constraints.
%This is all the more imperative given that the new DAS
%methodology is actually weaker than DP with a larger $\epsilon$,
%as it involves adjustments to preserve consistency 
%at several geographic levels (see Section~\ref{geo} above).

\subsection{{Transparency}}

One of the key claims when using differential privacy is that
``In turn, this allows an agency like the Census Bureau to quantify
the precise amount of statistical noise required to protect privacy.
This precision allows the Census to calibrate and allocate precise
amounts of statistical noise in a way that protects privacy while maintaining
the overall statistical validity of the data'' (\citeasnoun{abowd21}, p. 22).

 In fact,  this is true of any methodology, including swapping. It
is possible to select the swapping parameters to: (1)
include (more or less) records to be swapped, (2) the
attributes to be swapped, and (3) whether the swapping is
performed independently for each attribute. 

{
Releasing the $\epsilon$ parameter used in DP, as the Census Bureau does,
is certainly a step in the good direction. However, this alone does not make 
the protection methodology transparent. The postprocessing employed 
remains opaque to the users. One of the key criticisms against 
the swapping methodology employed until the Census 2010 was that
the swapping parameter (the proportion of swapped records) was not 
released to the public. But the Census Bureau did release
an upper bound for the proportion of swapped records. Given the simplicity
of swapping, this made the procedure pretty transparent. In addition,
swapping also assured that certain counts were preserved even at the 
block level, which afforded still greater transparency.}

{In our opinion, transparency is not just a matter
of parameter release; it also has to do with the complexity of the approach.
The more complex it is, the less transparent it is to the users. 
In this light, the current DP-based approach can be 
construed as being less transparent
than simple swapping.}

\section{Reconstruction and reidentification are different}
\label{linkage}

Reconstruction and reidentification are two different notions:
\begin{itemize}
\item Reconstruction is only the first step in the disclosure process.
Note that reconstructing from the outputs of statistical queries
or from released tabulations yields reconstructed data that include no 
identifiers.
\item Reidentification is a second and necessary step to complete
a disclosure attack. In this step, the reconstructed data are
{\em linked} to a particular individual. To this end, 
the attacker needs an external
data source that contains identifiers plus some attributes
that can be used to link with the reconstructed data. 
In the worst-case scenario (most favorable to the attacker),
the external data source may contain the {\em entire} original data
with identification information. For example, this worst-case scenario
makes sense if the attack is conducted by the same organization that protects
the data (in order to test the quality of reconstruction).
\end{itemize}

%A confusion between reconstruction 
%and reidentification is apparent in~\citeasnoun{garfinkelPETS}, where
%a U.S. Census Bureau methodologist claims a 
%database ``reconstruction and reidentification'' attack 
%on the 2010 Decennial Census with the following features:
\textcolor{black}{\citeasnoun{abowd21} (Appendix B)
and \citeasnoun{garfinkelPETS} describe the reconstruction
and reidentification as follows:}
\begin{itemize}
\item The microdata records of 308,745,538 people were ``reconstructed''.
\item Four external commercial databases of the 2010 US population
were used, which reported {\em Name}, {\em Address}, {\em Age}
and {\em Gender} of people.
\item The reconstructed records were linked to the commercial databases
to obtain a linked database with {\em Name}, {\em Address}, {\em Age},
{\em Gender}, {\em Ethnicity \& Race}. 45\% of records could be linked.
\item The linked database was compared to the U.S. Census Bureau
confidential data. It is claimed that the attack got all attributes
in 38\% of the linked records, or equivalently for 17\% of the U.S. population.
\item Hence, the authors claim reidentification of 17\% of the U.S. 
population, although he concedes that an outside attacker would not
know which reidentifications were correct. 
\end{itemize}

After that, the authors go on to criticize as flawed the protection system
used in the 2000 and 2010 Censuses, which relied on traditional SDC
techniques. This is used as a justification
to move to formal privacy, which amounts to DP. 
In~\citeasnoun{garfinkelPETS} it 
is explained that choosing the privacy budget $\epsilon$ is 
a public policy choice.

\subsection{Issues with reidentification claims}

There are several issues with the claims in~\citeasnoun{abowd21} (Appendix B):
\begin{enumerate}
%JOSEP2208. Rev. 5. Deleted.
%\item Reconstruction attacks in the literature {\em do not use
%external identified databases} and {\em do not assume the 
%attacker has access to the microdata.}
%The usual assumption in~\citeasnoun{dinur03} and even in~\citeasnoun{garfinkel19}
%(by the same author as~\citeasnoun{garfinkelPETS}) is that the attacker
%uses only the answers to statistical queries.
\item It is unclear what ``reconstructing'' 
the microdata of 308,745,538 people signifies. According 
to~\citeasnoun{vanriper20}, it amounted  
to re-generating microdata records from published
census block and tract tabulations, {\em i.e.}
from frequency tables with attributes {\em Census block},
{\em Age} and {\em Gender}). This is not true
reconstruction as DN describe. Note that, in general, the re-generation
of microdata from a frequency table is not unique,  
 because a frequency table contains less information
than the microdata it was computed from. 
%JOSEP2208. Rev. 1. Added.
{Hence, just re-generating one 
of the microdata sets that are compatible
with a certain frequency table does not qualify as reconstruction 
of the original data: in DN's notion of reconstruction, the accuracy
bounds are essential, and no such bounds are given for the Census 
so-called reconstruction~\cite{muralidhar2022}.} 
\item Reidentification means being able to link the records 
in anonymized microdata with the corresponding records in 
an external data set containing identifiers and covering a similar
population. This is not reconstruction. 
\textcolor{black}{Since} the attack was based on microdata re-generated
from frequency tabulations,  proper reidentification could 
only be conducted
from those cells with count 1. In all other cases, unequivocal 
reidentification is impossible.
\item It has been known at least since~\citeasnoun{sweeney00}
that matching a database containing demographic attributes such as
{\em Municipality\_of\_residence}, {\em Birthdate} and {\em Gender} against
an external database containing those same attributes plus 
identifiers for the 
same population is likely to yield a high proportion of reidentifications. 
In fact, \citeasnoun{ruggles21} show using a simulation
that most matches reported by the Census Bureau experiment
at a block level would
be expected randomly and {\em thus fail to demonstrate a credible
threat to confidentiality}\footnote{An alternative has also been 
proposed by~\citeasnoun{rugglespersonal} to investigate the impact of
reconstruction on reidentification. The idea is to first
match the external database to the reconstructed census data.
That yields a certain matching rate $r$. Then take those unmatched records from
the external database and compare 
them by block ID and PIK (the Protected
 Identification Key created by the Census Bureau for each original record) to
 the Census Edited File (the original confidential data). Let $r'$ be 
the reidentification rate resulting from this comparison.
If $r \approx r'$ then database reconstruction has little or no impact
on reidentification; to demonstrate that reconstruction 
increases the reidentification risk, $r$ should be substantially greater than
$r'$. The Census Bureau is yet to make this comparison.}.
%JOSEP. Citro comment.
Hence, the use of DP \textcolor{black}{may not be} necessary\footnote{
In fact, even if the
threat to confidentiality was credible, it is unclear that the 
Census's new DP-based TDA algorithm offers the best protection.
In~\citeasnoun{francis} it is shown that  
 race and ethnicity can be inferred with
more precision and less prior knowledge from TDA outputs than 
from the outputs of the Census previous protection algorithm.}.
\end{enumerate}

The above issues clearly show that, 
{rather than focusing on reidentification,
the Census experiment focuses on finding (non-unique) 
candidate reconstructions.} 
We show next that (mis)interpreting {reconstruction 
as reidentification} may in some situations 
overstate and in other situations understate the real risk of reidentification. 

\subsection{Misinterpreting reconstruction as reidentification 
may overstate or understate
the reidentification risk}

%We give here an illustration of how (mis)interpreting the 
%reconstruction accuracy as the risk of reidentification may 
%sometimes overstate the latter risk and sometimes understate it.

Recall that in the Census Bureau's ``reidentification'' procedure described
in~\citeasnoun{abowd21} and~\citeasnoun{garfinkelPETS}, and summarized at the beginning of 
Section~\ref{linkage},
reconstructed microdata reporting {\em Gender}, {\em Age},
{\em Race} and {\em Ethnicity} are linked to an external commercial
database reporting {\em Name}, {\em Address}, {\em Age} and {\em Gender}. 
Thus, linkage is performed using the {\em Age} and {\em Gender} attributes.
As a result of linkage, a linked database 
is obtained that reports 
{\em Name}, {\em Address}, {\em Age}, {\em Gender}, {\em Race}
and {\em Ethnicity}. 

Consider three scenarios at the block level:

\begin{itemize}
\item {\em Scenario 1.}
Block whose reconstructed data consist of 
10 individuals with {\em Age}=44, {\em Gender} = Male, 
 {\em Race} = White and {\em Ethnicity} = Not\_Hispanic.
The commercial database  
contains {\em Name}, {\em Address}, {\em Age}=44 and {\em Gender}=Male
for all individuals in this block.
\item {\em Scenario 2.}
Same data as in Scenario 1, but with an additional attibute
{\em Relationship}, which according to~\citeasnoun{garfinkelPETS}
is also collected for each person in a block and can take
17 different values. Assume that in, this scenario,
each of the 10 persons in the block has a different {\em Relationship}
value. Like in Scenario 1, 
the commercial database                              
contains {\em Name}, {\em Address}, {\em Age}=44 and {\em Gender}=Male
for all individuals in this block.
\item {\em Scenario 3.}
Block whose reconstructed data consist of
10 individuals with {\em Age}=44, {\em Gender} = Male,
such that such 
that all 10 of these individuals belong to different 
({\em Race}, {\em Ethnicity}) combinations.
The commercial database
contains {\em Name}, {\em Address}, {\em Race} and {\em Ethnicity} 
for all individuals in this block, but no {\em Age} or {\em Gender}.
%\item {\em Scenario 3.}
%Block whose reconstructed data consist of 
%10 individuals with {\em Age} = 44 and {\em Sex} = Male, such 
%that all 10 of these individuals belong to different 
%({\em Race}, {\em Ethnicity}) combinations.
%The commercial database
%contains {\em Name}, {\em Address}, {\em Age}=44 and {\em Sex}=Male
%for all individuals in this block.
%\item {\em Scenario 4.}
%Same data as in Scenario 3, but with the additional attibute
%{\em Relationship} taking a different value for each of the 
%persons in the block.
\end{itemize}

In Scenario 1, the Census Bureau's procedure 
would yield a 100\% reconstruction,
because the attacker would always be able to associate the correct
{\em Race} and {\em Ethnicity} to the 10 names
and addresses in the block.
Yet, claiming that this 100\% reconstruction amounts to 100\% reidentification  
is patently incorrect, because the attacker has no way to 
confirm the identification of
({\em Name}, {\em Address}) for the 10 individuals 
 who are indistinguishable from one 
another 
%JOSEP. Deleted. Citro comment.
% Claiming that reidentification has occurred because
%{\em Ethnicity} and {\em Race} have been correctly linked
%to a certain {\em Name} and {\em Address} is disingenuous
\textcolor{black}{--- ``reidentification'' in this case} 
can be attributed to the homogeneity
of a block and is not a true reidentification.
\textcolor{black}{The U.S. Census Bureau document~\citeasnoun{mckenna19} 
states that ``it is necessary to verify the proposed matches 
by comparing the suppressed identities in the microdata with 
the identities in the external dataset to see if the matches 
are true matches or false matches.''}

The above point that correctly reconstructing {\em Ethnicity} and {\em Race}
does not amount to reidentification becomes apparent in Scenario 2.
When the attribute {\em Relationship} is added with different
values for all 10 individuals, it becomes clear that the reidentification
probability for any specific individual is in fact $1/10$. 

%In Scenario 3, the Census Bureau's procedure would conclude that
%the ``reidentification'' probability is $1/10$ 
%(and hence the ``reidentification'' rate is 10\%), since the only way to 
%assign a particular individual based on ({\em Gender}, {\em Age}) 
%to ({\em Race}, {\em Ethnicity}) is to do it randomly. 
%But in fact every individual in the block is unique
%(in terms of ({\em Race}, {\em Ethnicity}) and uniquely
%corresponds to a particular {\em Name} and {\em Address}.
%Thus, the true reidentification rate could be as high as 100\%.  

In Scenario 3, both the probability of correct reconstruction 
and the probability of correct reidentification are 1, but for 
different reasons:
\begin{itemize}
\item Since all individuals have the same combination of 
({\em Age}, {\em Gender}), reconstructing the values of these attributes for the 
10 individuals is trivial, which yields 100\% 
reconstruction. {Note that if not all individuals
had the same combination, then the probability of correct
reconstruction would be less than 1.}
\item Since all individuals have different combinations 
of ({\em Race}, {\em Ethnicity}), unequivocally linking each of the 10 records
in the reconstructed data to its corresponding record in the commercial
database is straightforward, which yields 100\% reidentification.
\end{itemize}

The above shows that reconstruction 
and reidentification are different notions. 
The bottom line is as follows: whereas {\em reconstruction is helped by
homogeneity of the missing confidential attributes,
reidentification is helped by heterogeneity of the
quasi-identifiers through which linkage is performed.}

Hence, the reconstruction procedure described 
in~\citeasnoun{abowd21} and~\citeasnoun{garfinkelPETS}
does not yield an appropriate measure of reidentification risk. 
In fact, it is likely to overstate the reidentification 
risk (as in Scenario 1), since at the block
level ({\em Race}, {\em Ethnicity}) can be expected to be fairly
homogeneous, which makes Scenario 1 more likely than 
Scenario 3~\cite{ruggles21}.

%JOSEP2208. Rev. 1 and 2. I replace "Ironically" by "Interestingly"
{Interestingly}, researchers at the U.S. Census Bureau have 
performed in the past extensive research in reidentification
risk ({\em e.g.},~\citeasnoun{winkler}). To assess the true risk
of reidentification, 
it is necessary to assume the following. At the block level, 
the reconstructed data consist of 
({\em Block\_ID}, {\em Gender}, {\em Age}, {\em Race}, 
{\em Ethnicity}, {\em Relationship}) and the attacker has the attributes 
({\em Name}, {\em Address}, {\em Block\_ID}, {\em Gender}, {\em Age}, 
{\em Race}, {\em Ethnicity}, {\em Relationship}). 
The objective of reidentification is to uniquely link a record from 
the (unidentified) reconstructed data to a record in the (identified) 
attacker's data thereby attaching ({\em Name}, {\em Address}) to 
the reconstructed data. {\em Such a procedure will correctly assess 
the reidentification risk in the scenarios described above}; as mentioned
above and in~\citeasnoun{mckenna19}, once the linkage is established, reidentification
needs to be validated by checking 
that the linkage is unique and that identities (name and address in this case) match between 
the attacker's record and the original record to which the unidentified 
reconstructed record corresponds.
%To reidentify a record in the unidentified data,  
%it must be uniquely linkable to another record in the attacker's data.
{Reconstructing unidentified records, in itself, does not 
pose a real disclosure threat. ``Reconstruction'' in the DN sense also
requires to be supplemented by correct re-identification. Only then does it constitute
real disclosure.}

\section{Conclusions and future work}
\label{conclusion}

In this paper, we have reassessed the feasibility 
of reconstructing a data set based
on the outputs of statistical queries computed on it.
The danger of reconstruction has been cited
as an argument to justify 
the use of differential privacy in official statistics,
most notably in the case of the 2020 Census of the U.S.A.
Using DP, however, will most likely result in a decrease
of the utility of the statistical outputs of that Census.
% Deleted Citro.
% which 
%has led to some states engaging in lawsuits against
%the U.S. Federal Government.
This paper has investigated to what extent reconstruction
is a real danger. 

We first examined the state of the art in reconstruction theory 
---Dinur and Nissim's framework--- and  
we concluded that \textcolor{black}{local} 
or input protection appear as good ways to resist reconstruction.
If the U.S. Census Bureau were to stick to the so-called CD-model
and produced \textcolor{black}{locally protected 
or input-protected data} ({\em e.g.}, using RR or microdata masking discussed
in~\citeasnoun{hundepool}, or the methods used in the 2010 Census), then
reconstruction would not be a real danger: at most the 
attacker would be able to reconstruct  
 the \textcolor{black}{locally protected} 
or the input-protected data, 
rather than the original data.
%Then, we have explored the options in case output 
%protection is used to compute queries on an $n$-record
%original data set. According to Dinur and Nissim, 
%$O(\sqrt{n})$ noise is needed to thwart
%a polynomial adversary who is allowed to submit $O(n)$ queries. 
%Differential privacy 
%actually adds $O(n)$ noise in such a situation, due to sequential composition,
% which is more noise than necessary, thereby leading
%to unnecessary utility loss.  
Differential privacy is also an option,
but it may add more noise than strictly required to counter reconstruction, 
thereby leading to unnecessary utility loss~\cite{ons,hotz,bach}, or it may offer less
protection than previous approaches~\cite{francis}. 

%JOSEP2208. Rev. 1. Deleted
%We have subsequently revisited a case study 
%based on data from~\citeasnoun{garfinkel19}. Unlike the authors
%of that paper, we conclude 
%that {\em even} for a very small data set, {\em even} 
%when a substantial number of statistics are released, 
%{\em reconstructing the data set is practically impossible
%if simple SDC techniques are properly applied.}
%The alternative of using DP in this case study 
%may also protect against reconstruction,
%but it does so at the cost of yielding worthless statistics.
%
%It is important to realize that the data set and the statistics 
%from~\citeasnoun{garfinkel19} are
%a fictional but {\em plausible} scenario created by senior
%methodologists at the U.S. Census Bureau. If this is the 
%scariest scenario they can come up with, there 
%seems to be little to worry about database reconstruction. 
%Yet the SAT solver techniques they employed 
%might be usefully leveraged as a risk assessment tool to 
%enhance the protection process.

We then highlighted the relevance for protection 
of the geographic levels
at which exact population counts are preserved.
No matter the SDC methods used, preserving
counts in small geographies facilitates reconstruction, while 
not preserving counts in small geographies goes
a long way towards avoiding reconstruction, but also reducing utility.
%This is an important
%observation because the new Census Bureau's DAS methodology (based on DP) 
% involves postprocessing adjustments to preserve
%population counts at several geographic levels.
 
Finally, we have warned against using the amount of reconstruction
as a measure of reidentification risk, which 
results in exaggerated reidentification risk.
 Whereas reconstruction requires only query outputs 
or tabular unidentified outputs
and is favored by the homogeneity of the missing attribute values, 
reidentification also needs 
external identified sources and is favored by the heterogeneity of 
the values of the attributes used for linking with those sources.

%In summary, the 'sky is falling' panic
%of database reconstruction resulting in mass disclosure 
%is unwarranted. The 
%Moreover, proper use of existing traditional SDC techniques
%seems a better option than differential privacy 
%to deliver useful statistical information while 
%protecting against reconstruction attacks. 

\textcolor{black}{An additional concern are the successive
increases of the value of $\epsilon$ during the process. Increasing
$\epsilon$ implies a loss of privacy.
The Census Bureau started with Laplace noise addition with $\epsilon=4.5$ 
and subsequently increased to $\epsilon=10.2$. 
To further improve the utility of the released data, the Bureau
adopted zero-concentrated DP (with noise from a discrete
Gaussian distribution,~\cite{bun16}) in place of Laplace noise. 
The parameter $\rho$ of zero-concentrated DP can be used 
to compute equivalent values $\epsilon$ and $\delta$ 
for $(\epsilon,\delta)$-DP. For a given 
$\rho$ there are many equivalent combinations $(\epsilon,\delta)$.
However, for fixed $\delta$ (in the case of the Census 
it is $\delta=10^{-10}$), then each $\rho$ has a single
equivalent $\epsilon$.
In 2021, this equivalent
global $\epsilon$ was 19.61~\cite{census21}. This value was further  
revised to a global $\epsilon=39.907$ in year 
2022~\cite{census22}. This has a great impact on privacy.
The privacy level associated with 
$\epsilon=39.907$ is worse than the privacy level given by 
$\epsilon=4.5$ by a factor $e^{39.907}/e^{4.5} = 2.382 \times 10^{15}$.}
Referring to Apple's use of $\epsilon=14$, 
Frank McSherry, one of the inventors of DP, commented that it was 
``pointless'' in terms of privacy~\cite{wired}. 
Actually, \textcolor{black}{$\epsilon=39.907$ is over $1.78 \times 10^{11}$ 
times worse than $\epsilon=14$.} 
In~\citeasnoun{abowdadd21},
the \textcolor{black}{then} Census's Chief Scientist said about $\epsilon$ that 
``specifically it limits the statistical power of all possible tests 
for whether a particular individual's data record (or portions thereof) 
was used to produce a collection of statistics versus the record of another, 
arbitrary individual.'' 
\textcolor{black}{With $\epsilon=39.907$ used in 2022, this implies that 
the probability that a particular individual's data record 
was used can be over $2.14 \times 10^{17}$ times higher 
versus the record of another, arbitrary individual. Since 
the current U. S. population is only $3.31 \times 10^8$, 
 with an $\epsilon=39.907$ any target 
U.S. inhabitant might be reidentifiable. In fact, this also held
for the $\epsilon=19.61$ used in 2021.}

\textcolor{black}{
Even with this relaxation of the value of 
$\epsilon$, there are still very serious utility concerns. Consider the following report in the 
New York Times~\cite{nytimes}: 
``According to the 2020 census, 14 people live there [in Census Block 1002 in downtown Chicago] --— 13 adults and one child. Also according 
to the 2020 census, they live underwater. Because the block consists entirely of a 700-foot bend in 
the Chicago River.'' Or this analysis from Cornell University~\cite{cornell}
%\footnote{\url{https://pad.human.cornell.edu/census2020/index.cfm#das}} 
of the 2021 Census DAS release for New York state which shows that in 6.1\% of the blocks, 
the household population is greater than 0, but the number of occupied houses is 0; in 2.5\% of the blocks, 
the household population is less than the number of occupied houses (which means there is less than 1 person
per household); and in 0.8\% of the blocks, the household population is 0, but the number 
of occupied houses is greater than 0. These results are impossible and would not have occurred in the 2010 Census. 
Thus, even with large $\epsilon$, the differentially private noise addition procedure 
is not capable of providing accurate and consistent results. 
In fact, the US Census Bureau recently announced that ``for the time being, the ACS PUMS [American Community Survey Public Use Microdata Sample] data product will still 
be protected using traditional disclosure avoidance methods'', since it is ``not clear that differential 
privacy would ultimately be the best option.''\cite{census22b} }

\textcolor{black}{In this study, we conclude that} 
the concern of database reconstruction resulting in mass disclosure is unwarranted. 
We believe that these claims are based on a comparison that is incomplete and opaque --—only the Census Bureau can assess or verify the true reidentification results. 
\textcolor{black}{Other researchers, some of them mentioned above, have raised serious concerns 
regarding the accuracy and consistency of the output. Hence, it is not clear that differential privacy is the best option for the 2020 decennial census data. We suggest a comprehensive, independent, fully documented, peer-reviewed assessment of the efficacy of alternative methods.}

%JOSEP2208. Rev. 1. Removed appendix.

\end{document}